\documentclass[apj,twocolappendix]{emulateapj}
\usepackage{ulem}
\PassOptionsToPackage{hyphens}{url}\usepackage{hyperref}
\usepackage{natbib}
\usepackage[]{amsmath}
\usepackage[]{amsthm}
\newcommand{\bjdtdb}{BJD$_{\textrm{TDB}}$}

\newcommand{\zmeas}{\ensuremath{z_{\rm meas}}}

\newcommand{\numeas}{\ensuremath{\nu_{\rm meas}}}
\newcommand{\lmeas}{\ensuremath{\lambda_{\rm meas}}}
\newcommand{\bearth}{\ensuremath{\vec{\beta_{\Earth}}}}
\newcommand{\bstar}{\ensuremath{\vec{\beta_*}}}
\newcommand{\gearth}{\ensuremath{\gamma_{\Earth}}}
\newcommand{\gstar}{\ensuremath{\gamma_*}}

\newcommand{\zgr}{\ensuremath{z_{\rm GR}}}

\shorttitle{Barycentric Corrections at 1 cm/s}
\shortauthors{Wright \& Eastman}

\begin{document}
\title{Barycentric Corrections at 1 cm/s for precise Doppler velocities}
\author{J.~T.~Wright\altaffilmark{1,2},J.~D.~Eastman\altaffilmark{3,4}}

\altaffiltext{1}
{Center for Exoplanets and Habitable Worlds,
525 Davey Lab,
The Pennsylvania State University,
University Park, PA, 16802}

\altaffiltext{2}
{Department of Astronomy \& Astrophysics,
525 Davey Lab,
The Pennsylvania State University,
University Park, PA, 16802}

\altaffiltext{3}
{Las Cumbres Observatory Global Telescope,
6740 Cortona Dr,
Goleta, CA, 93117, USA}

\altaffiltext{4}
{Department of Physics, Broida Hall,
University of California Santa Barbara,
Santa Barbara, CA, 93117, USA
}

\begin{abstract}
The goal of this paper is to establish the requirements of a barycentric correction with an RMS of $\lesssim 1$ cm/s, which is an order of magnitude better than necessary for the Doppler detection of true Earth analogs ($\sim9$ cm/s). We describe the theory and implementation of accounting for the effects on precise Doppler measurements of motion of the telescope through space, primarily from rotational and orbital motion of the Earth, and the motion of the solar system with respect to target star (i.e.\ the ``barycentric correction'').  We describe the minimal algorithm necessary to accomplish this and how it differs from a na\"ive subtraction of velocities (i.e.\ a Galilean transformation).  We demonstrate the validity of code we have developed from the California Planet Survey code via comparison with the pulsar timing package, TEMPO2.  We estimate the magnitude of various terms and effects, including relativistic effects, and the errors associated with incomplete knowledge of telescope position, timing, and stellar position and motion.  We note that chromatic aberration will create uncertainties in the time of observation,
which will complicate efforts to detect true Earth analogs.  Our code is available for public use and validation.
\end{abstract}

\section{Purpose and Plan}

The purpose of this document is to establish the minimal requirements of a barycentric correction to Doppler velocimetry that is valid at the 1 m/s, 10 cm/s, and 1 cm/s levels (which are an order of magnitude better than necessary for the Doppler detection of gas giant planets, close-in terrestrial planets, and true Earth analogs, respectively) and to explicate the mathematics of its calculation. The problem of precise timing of astrophysical phenomena has been solved to a precision more than necessary for such work.  For instance, \citet{stumpff79} presented an early implementation of a barycentric correction for precise Doppler measurements, and \citet{kopeikin99} present a comprehensive, post-Newtonian solution to the general problem of Doppler shifts of binary stars from their barycentric frame to that of a terrestrial observatory. The pulsar community employs the TEMPO2 software \citep{hobbs06, edwards06} which uses a fully general-relativistic framework to predict and analyze of pulsar pulse arrival times.

The problem of Doppler exoplanet detection is a restricted version of this general problem, and so does not require most of the machinery of general relativity or a full relativistic modeling of the exoplanetary system.  Specifically, the astrometric displacements and space velocities of stars due to planets are small enough that they can be treated as linear perturbations or ignored for many purposes, and at any rate, the motions and inhomogeneities of stellar atmospheres \citep[i.e., jitter, e.g.][]{wright05} limit the precision of Doppler work to levels that do not require the rigor provided by, for instance, TEMPO2.  Finally, precise stellar Doppler velocities are measured with respect to some fiducial spectrum or wavelength scale, not with respect to an absolute standard such as an atomic clock, and this necessitates specialized formulations of the relativistic Doppler equations.

In this manuscript, we define the ``barycentric correction'' of a star as the transformations of measured Doppler shifts (with respect to a fiducial spectrum or wavelength scale) to the ``true'' redshifts that would be measured from a platform stationary with respect to the stellar system barycenter.  That is, what would be measured from the Solar System Barycenter (SSB) if the barycenter of the target stellar system system had no proper motion.  In our formulation, all Doppler shifts due to the Earth's motion with respect to the stellar barycenter, (including perspective effects, secular acceleration, and parallax-proper motion cross terms) are ``nuisance terms'' to be removed by the generalized barycentric correction procedure.

In \S \ref{sec:techniques}, we summarize the two primary precise stellar Doppler velocimetric techniques and how they relate to the barycentric correction.  In \S \ref{sec:relativisticdoppler}, we give a didactic presentation of relativistic Doppler shifts, and in \S \ref{sec:realstars}, we apply the terminology we have developed to the practical case of barycentric correction in real stars.  Section \ref{sec:barycorr} describes \verb#BARYCORR#, our public IDL code that implements this correction, and how it compares to TEMPO2.

In \S \ref{sec:magnitude}, we discuss the magnitude of the various terms in our expression for barycentric correction, which will allow practitioners to construct a thorough error budget given uncertainties in the various inputs to a barycentric correction algorithm, such as observatory position, stellar position and space motion, and timing of the observations. Finally,  Appendix \ref{sec:tempo} describes how TEMPO2 may be used to include effects that we have ignored for a higher precision correction.

Table~\ref{variables} provides a definition of the symbols used in this manuscript.  The symbols and equations required for the practical computation of barycentric corrections are listed first.  Symbols only used in the explanatory and didactic sections are listed separately.

\begin{deluxetable*}{clc}
\tablewidth{7.0in}
\tablecolumns{3}
\tabletypesize{\scriptsize}
\tablecaption{Variables and symbols used in this manuscript \label{variables}}
\tablehead{\colhead{Symbol} & \colhead{Meaning} & \colhead{Example equation}}
\startdata
\multicolumn{3}{c}{Essential symbols for barycentric motion correction}\\
\bearth & Total velocity of observatory in SSB frame in units of $c$ & \ref{elementaryppm}, \ref{BC} \\
$\vec{\beta_s}$ & Velocity of the barycenter of the stellar system (assumed constant) in the SSB frame in units of $c$ & \ref{bstar}, \ref{betas}, \ref{BC}\\
$c$ & Speed of light & \\
\gearth & $1/\sqrt{1 - \bearth^2}$& \ref{BC} \\
$\vec{\mu}$ & Proper motion vector of star in units of radians per time at epoch of observation & \ref{BC} \\
$\vec{\rho}$, $\hat{\rho}$, $\rho$ & Vector, unit vector, and distance from observatory to star & \ref{vecx}, \ref{BC} \\
$\vec{r}_0$,$\hat{r}_0$,$r_0$, & Vector, unit vector, and distance to star from SSB at epoch of coordinates & \ref{betas}, \ref{vecr}, \ref{BC} \\
$v_r$ & Radial velocity of barycenter of stellar system at nominal epoch & \ref{vecr} \\
$\vec{x}$ & Vector from SSB to observatory with units of length & \ref{vecx} \\
$z_B$ & Total redshift due to all effects of barycentric motion & \ref{zb}, \ref{BC} \\
$\zgr$ & Redshift due to gravitational time dilation due to Solar System objects & \ref{GR},\ref{zb} \\
\zmeas & Redshift measured by a spectrograph & \ref{zb}\\
$z_{\rm true}$& Barycentric-motion-corrected redshift. $c z_{\rm true}$ is the ``radial velocity'' of the star to report & \ref{ztrue}, \ref{zb} \\
\hline
\multicolumn{3}{c}{Other symbols used in this manuscript}\\
$\prime$ & Superscript indicating quantity at the time and location of the template/reference observation & \ref{nuprime}\\
$\hat{}$ & Indicates a unit vector & \\
$*$ & Subscript indicating the source (i.e.\ the star) & \\
$\earth$ & Subscript indicating the observer (i.e.\ the observatory in the SSB frame) & \\
0 & Subscript indicating an arbitrary, reference redshift measurement & \ref{i0} \\
$i$ & Subscript indicating an epoch measurement (as contrasted with the reference measurement) & \ref{i0} \\
emit & Subscript indicating idealized emitted $\lambda$ or $\nu$ & \ref{zdef} \\
meas & Subscript indicating measured quantity & \ref{zdef}, \ref{BC}\\
$\hat{\alpha}, \hat{\delta}, \hat{u}$ & unit vectors in RA, Dec, and North Celestial Pole directions used to calculate $\vec{\mu}$ & \ref{coords}\\
$\vec{\beta_p}$ & Velocity of the star in the frame of the stellar system barycenter (i.e.\ planetary perturbations) & \ref{bstar}\\
$\Delta \vec{\beta_p} $ & Change in motion of star due to planets between observations $i$ and 0 & \ref{i0}\\
$\Delta t$ & Time since epoch of coordinates of star & \ref{vecr}, \ref{lt}\\
$d$ & Distance from Earth to a solar system object & \ref{GR} \\
$\gamma$ & $1/\sqrt{1 - \beta^2}$ & \\
$G$ & The gravitational constant & \ref{GR}, \ref{extra}\\
$\lambda$ & Wavelength & \ref{zdef} \\
$M_i$, $M_\odot$ & Mass of a solar system object or the Sun & \ref{GR}, \ref{extra} \\
$\mu$, $\mu_\alpha$, $\mu_\delta$ & magnitudes of $\vec{\mu}$ (total and in the RA and Dec directions ($\dot{\alpha}\cos\delta$ and $\dot{\delta}$) & \ref{coords}\\
$\nu$ & Frequency & \ref{zdef} \\
$\nu^\prime$ & Frequency of a spectral feature measured at time and location of template & \ref{nuprime} \\
$\varpi$ & Parallax angle & \\
$\pi$ & The numerical constant & \\
$z$ & Redshift & \ref{zdef} \\
$z^\prime$ & Redshift relating template or reference wavelengths to emitted wavelengths & \ref{nuprime}\\
$z_{\rm LT} $ & Redshift due to light travel time effects from space motion & \ref{lt} \\
$z_{\rm SD}$ & Redshift due to the Shapiro delay & \ref{extra} \\
$z_{\rm GR*}$ & Redshift due to gravitational time dilation due to objects in the stellar system & \ref{GR},\ref{zb} \\
\hline
\enddata
\end{deluxetable*}
\section{Summary of the techniques}
\label{sec:techniques}
In both primary precise Doppler techniques, the position of stellar spectral features on a detector is associated with a combination of instrument changes (the problem of precise wavelength calibration) and astrophysical redshift $z$ from some fiducial position, the latter resulting from a combination of the motions of the observatory and target star.

\subsection{Absorption Cell Calibration}

In the absorption cell method, a stellar spectrum is passed through a gas, usually iodine, which imprints an absorption spectrum of precisely known form on the starlight.  This allows for a precise wavelength scale to be established for a high signal-to-noise ratio stellar spectrum.  A ``template'' observation is made of the star to establish the true stellar spectrum; this spectrum is usually taken without the absorption cell, and so has a less-precisely known wavelength scale.  Relative Doppler shifts are measured from the wavelength shift of spectral features in the rough template solution (at wavelength $\lambda^\prime$) and a measured spectrum (at wavelength $\lmeas$), $(1+\zmeas)=\lmeas/\lambda^\prime$.  This is usually accomplished through a forward modeling of the measured spectrum \citep{butler96}.

\subsection{Emission Line Calibration}
The other primary method of precise velocimetry is to establish an accurate wavelength scale for every observation with a standard such as a ThAr lamp or laser comb, and to minimize the effects of instrument changes through stabilization of the spectrograph.  The primary procedural difference from the iodine technique is that the wavelength solution is established by a beam that is not coincident with the starlight; typically the calibration light (containing sharp spectral features of precisely known frequencies) is sent through a separate fiber and imaged simultaneously with the starlight, or else exposures of such calibration light ``brackets'' the stellar observation.

In this method, shifts are usually measured with a cross-correlation technique \citep{baranne96, pepe02}, although occasionally forward-modeling is used \citep{anglada12}. A mask or template is devised (often a binary mask tailored to the particular spectral type of the star) and the cross-correlation function (CCF) of the mask with the observed spectrum determines the shift of a spectral feature from a fiducial wavelength.  The wavelength solution then provides the translation of this shift into a measured Doppler shift.

\subsection{Application to Barycentric Corrections}

In both techniques, redshifts are measured as the shift of spectral features with respect to some fiducial wavelength.  In the emission line technique, this fiducial can be chosen such that the measured redshifts correspond closely to the true center-of-mass velocity with respect to the observatory (the ``absolute'' velocity of the star).  In the absorption cell technique, the fiducial is typically the observed wavelength of the spectral feature at the time of the template observation.

The barycentric correction procedure we describe here applies equally to both techniques, with \zmeas\ being the observed redshift with respect to the fiducial wavelength.

\section{Relativistic Doppler Shifts for Stars and the Earth}
\label{sec:relativisticdoppler}

\subsection{Elementary Forms}

\label{sec:elementary}

So that we may unambiguously define our variables, here we provide some elementary forms for special relativistic Doppler shifts.  Except where otherwise specified, we perform all calculations and define all terms in the inertial frame of the SSB (i.e.\ the center of mass of the Solar System) and we normalize all velocities $\beta$ by the speed of light.

The redshift $z$ is defined:

\begin{equation}
\label{zdef}
z \equiv  \frac{\lambda_{\rm meas} - \lambda_{\rm emit}}{\lambda_{\rm emit}} = \frac{\nu_{\rm emit}-\numeas}{\numeas},
\end{equation}
\noindent where $\nu_{\rm emit}$ is the emitted frequency and $\numeas$ is the observed frequency.

Relativistically, we can write, for a moving {\em observer} on Earth (with velocity \bearth), a {\em stationary} source at infinity (and ignoring gravitational effects).

\[
\label{numeas}
\numeas = \nu_{\rm emit} \gearth (1+\bearth\cdot\hat{r}),
\]
\noindent where $\gearth = 1/\sqrt{1-\bearth^2}$.  So,
\[
\zmeas + 1 = \frac{1}{\gearth (1+\bearth\cdot\hat{r})}.
\]

\noindent Here, $\hat{r}$ points in the opposite direction of the light's travel.  This means that for a moving observer, going away from the source, we get a positive (redshift) contribution from the speed of the observer.  Next, note that the gamma factor makes the photons appear {\it blueshifted} because, in the frame of the SSB, the Earth's clocks run slow because the Earth is moving quickly, so Earth observes more wavecrests per clock tick than it would at rest with respect to the SSB.\footnote{It may at first seem to violate the relativity of velocities that the moving observer sees blueshifted photons from a term that, at first blush, resembles the term that generates the transverse Doppler shift, which is always a redshift.  The resolution of this paradox involves shifting into the Earth's frame and seeing that what appears to be a transverse Doppler shift actually contains a radial component due to the aberration of starlight (a real effect in the frame of the Earth), and that the Earth's frame is not inertial over an entire orbit.}

For a moving {\em source} at infinity (moving with velocity \bstar), and a {\em stationary} observer, relativistically (and ignoring gravitational effects) we write:

\[
\label{Dop}
\numeas = \frac{\nu_{\rm emit}}{\gstar(1+\bstar\cdot\hat{r})},
\]
\noindent so
\[
\zmeas + 1 = \gstar (1+\bstar\cdot\hat{r}).
\]

\begin{figure}
\includegraphics{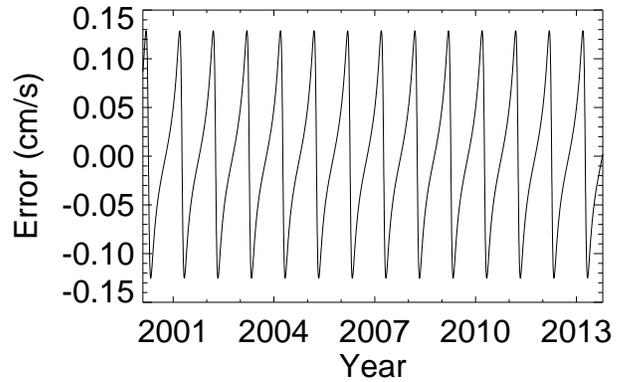}
\caption{The Shapiro delay for $\tau$ Ceti over 14 years, showing at most a 1.5 mm/s effect, maximized when the Sun is only 23$^\circ$ away from $\tau$ Ceti. In practice, ground based optical astronomers cannot observe targets much closer than 32$^\circ$ from the Sun (e.g., at nautical twilight and airmass=3), so this is a worst-case value.
}
\label{fig:shapiro}
\end{figure}

A proper formulation that takes the effects of general relativity into account requires much more rigor than we present here.  Fortunately, there are only two factors that might be important at the 1 cm/s level:  gravitational blueshift due to the Sun and the Earth (only the former is time variable), and the Shapiro delay \citep{shapiro64}.  The latter occurs because light passing near a gravitational object has a longer path to take to the telescope due to gravitational deflection, and the variation in this path length as the light path sweeps by the object appears as a frequency variation. The effect for $\tau$ Ceti is at most 1.5 mm/s, and is maximized when the Sun is only 23$^\circ$ away from $\tau$ Ceti (see Figure \ref{fig:shapiro}). In practice, ground based optical astronomers cannot observe targets much closer than 32$^\circ$ from the Sun (e.g., at nautical twilight and airmass=3), so the effect will be smaller still. The effect for all other masses in the Solar System is negligible; for a star passing within 1$^\prime$ of Jupiter, it is $\ll 1$mm/s.

Photons from the source star will gain energy as they enter the Solar System's potential well, and so become blueshifted by:
\begin{equation}
\label{GR}
z_{\rm GR} \approx  -\sum_i\frac{GM_i}{c^2d_i},
\end{equation}

\noindent where $G$ is the gravitational constant.  The sum on $i$ in Eq.~\ref{GR} is over all bodies in the Solar System (only the Sun and Earth are important at 1 cm/s) with masses $M_i$ and distances from the observatory $d_i$.  Typical values near the Earth are of order 3 m/s, and vary with amplitude $\sim 3$ cm/s.  There is an equivalent term for photons leaving the stellar system, $(1+z_{\rm GR*})$, which is constant for most stars (since the dominant term is the star itself, and for non-pulsating stars, photons are emitted at constant distance from its center).

In the SSB frame, then, the frequency of photons received by a moving observer from a moving source (ignoring the Shapiro delay) is then

\begin{equation}
\label{elementary}
\numeas = \nu_{\rm emit}\frac{(1+z_{\rm GR*})}{(1+\zgr)}\frac{\gearth (1+\bearth\cdot\hat{r})}{\gstar(1+\bstar\cdot\hat{r})}.
\end{equation}

\subsection{Philosophy of ``barycentric correction''}

The velocity of a star can be expressed as

\begin{equation}
\label{bstar}
\bstar = \vec{\beta}_s + \vec{\beta}_p,
\end{equation}

\noindent where $\vec{\beta}_s$ represents the systemic velocity of the star, i.e., the constant motion of the barycenter of its system, and $\vec{\beta}_p$ represents the perturbative (or ``planetary'') velocities, i.e.\ the combined effects of orbital companions.  We can further define $v_r = c\vec{\beta}_s \cdot \hat{r}_0$, the constant systemic radial velocity of the stellar system at some nominal epoch (at which $\hat{r}_0$ is defined), about which we measure planetary perturbations.

Strictly speaking, this addition should be performed relativistically; that is, $\vec{\beta}_p$ is defined in the inertial frame of the star, but $\vec{\beta}_s$ is defined in the frame of the SSB. Expanding the formula for the relativistic addition of velocities, we can see that the corrections are of order $\beta_p \beta_s^2\sim 10^{-14}$, and so not relevant even at the 1 cm/s level.

When the radial velocity of a star is measured, it is compared with some fiducial and, over time, a series of velocities produces a radial velocity (RV) time series.  This time series is used to infer the existence of orbiting companions, stellar oscillations, starspots, and other sources of radial velocity variation.  A fully rigorous solution to the problem would be to model the entire system, including the motion of the star, its companions, and the motion of the Earth, and fit this model to the data, as is done with pulsar timing models.  The model could then include orbital companions, such as planets, as model parameters.

An alternative approach, most commonly used in planet detection, is to remove all effects not related to planetary perturbations from the radial velocity data, so that one's model may be simpler and treat only the planets themselves, plus some constant RV offset.  The most obvious of the effects to remove is the motion of the Earth around the Sun, but there are also relativistic and perspective effects that will appear to create RV variation even when the star is, in fact, stable.

An additional complication is that one is not measuring absolute velocities of the star, so one is not truly measuring frequencies with respect to the emitted frequency $\nu_{\rm emit}$, but with respect to some other, fiducial frequency, $\nu^\prime$.  In the iodine technique, this frequency is $\numeas(t_{\rm template})$, and in the emission line technique, it is often the frequency of the spectral feature in the solar or model spectrum that generated the CCF mask.  Regardless, this reference frequency $\nu^\prime$ is related to $\nu_{\rm emit}$ by some redshift.  For an algebraic convenience that will become clear presently, we will define this redshift $z^\prime$ (which is constant for all spectral features) as the value that satisfies the relation:

\begin{equation}
\label{nuprime}
\nu^\prime = \frac{\nu_{\rm emit} (1+z_{\rm GR*})}{\gstar (1+z^\prime)}.
\end{equation}

We next define $z_{\rm true}$ as the redshift that a fiducial, stationary observer would measure in the frame of the SSB in this constant $\hat{r}_0$ direction comparing measured features to a given reference frequency $\nu^\prime$:

\begin{equation}
\nu_{\rm meas,true} = \frac{\nu_{\rm emit} (1+z_{\rm GR*})}{\gstar(1+\bstar\cdot\hat{r}_0)},
\end{equation}

\begin{equation}
(1+ z_{\rm true}) =  \frac{\nu^\prime}{\nu_{\rm meas,true}} = \frac {(1+(\vec{\beta}_s +\vec{\beta}_p)\cdot\hat{r}_0)}{(1+z^\prime)} \label{ztrue}.
\end{equation}

\noindent This is the frequency one would measure in a ``true''ly radial direction (i.e., a constant one).  Since the star moves with respenoy yrh ct to the SSB, we must imagine that our fiducial observer is actually a set of observers at rest in the SSB, each making their observation at the moment that the star lies in the direction $\hat{r}_0$.  If such observers made a series of measurements $z_{\rm true,i}$  (each relating $\nu_{\rm meas,i}$ to $\nu^\prime$), then the difference in these measured redshifts from a reference measurement $z_{\rm true,0}$ would relate to the changes in the star's motion due to planetary companions by

\begin{equation}
\label{i0}
z_{\rm true,i} - z_{\rm true,0} = \frac{\Delta\vec{\beta}_p \cdot \hat{r}_0}{(1+z^\prime)}.
\end{equation}

\noindent That is, the measured differences would be proportional to the actual changes in the line-of-sight velocity of the star, and the constant of proportionality would differ from unity by some very small amount, which could be neglected entirely for low-amplitude systems, or estimated for high amplitude systems with even modest knowledge of $z^\prime$.  Equivalently, the observer could shift the wavelength scale of the model spectrum so that $z^\prime$ has its ideal value of zero.

The ``barycentric correction'', then, is a function or procedure that reduces measured redshifts to a series of values of $z_{\rm true}$:

\begin{equation}
{\rm bary}(z_{\rm meas,i}) = z_{\rm true,i} = \frac{\Delta\vec{\beta}_p \cdot \hat{r}_0}{(1+z^\prime)} + {\rm bary}(z_{\rm meas,0}).
\end{equation}

\noindent These true redshifts can then trivially be interpreted as changes in the projection of $\vec{\beta}_p$ in the radial direction; that is, the radial velocity of the star is $c z_{\rm true}$.

This is not strictly possible to measure; at the very least the proper motion of the star makes the ``radial direction'' in the frame of the stellar system variable, creating perspective effects on the star's reflex motion. But for planetary motion, the perturbative velocities are small (often at the edge of detection). At 1 cm/s precision, only the largest planetary perturbations will show evidence of perspective effects, which are of order $\mu \Delta t \beta_p$ (so at worst, over a decade, $\mu \sim 10^{-5}$ rad/yr, $\Delta t \sim 10$ yr, $c\beta_p \sim $ 300 m/s, yielding $\sim$ 3 cm/s).   Complex systems, such as binary star systems with planetary companions, may not be perfectly amenable to the ``barycentric correction'' prescription we offer here, and so may require a more rigorous treatment.

In the limit that these perturbations are small, however, we can define $z_B$ as the redshift one must correct measured redshifts by to obtain the ``true'' redshift one seeks from which to derive the true center-of-mass motion of the system in question, independent of the motion of the Earth:

\begin{equation}
\label{zb}
{\rm bary}(\zmeas) = (1+\zmeas)(1+z_B) - 1 = z_{\rm true}.
\end{equation}

Note the correction to \zmeas\ is multiplicative, not additive.  One should not, therefore, formulate the ``barycentric correction'' as the velocity that should be added or subtracted from \zmeas\ to find $z_{\rm true}$, as one often does for work at coarser precisions.  That is, the difference between \zmeas\ and $z_{\rm true}$ includes a cross term of magnitude 3 m/s:

\begin{equation}
z_{\rm true}- \zmeas = z_B +z_B \zmeas,
\end{equation}

\noindent which requires knowledge of \zmeas \ and cannot be known in advance to arbitrary precision.  It is practically desirable to formulate barycentric corrections such that one may precompute a universally-agreed upon quantity (here, $z_B$) which is then applied (via Eq.~\ref{zb}) to measured velocities to perform the barycentric correction.

\section{Barycentric correction for real stars}
\label{sec:realstars}

Light was originally emitted from the star at frequency $\nu_{emit}$ in its own frame, but due to effects on its way out of that stellar system, the shift to the SSB frame, gravitational effects in the solar system, and the motion of the telescope with respect to the SSB, it will actually be measured at the telescope to have frequency $\nu_{meas}$.
\subsection{Stellar Positions and the Systemic Radial Velocity}
\label{sec:stellarmotion}
\subsubsection{Proper motion}

\label{lighttravel}
The star sits at (retarded) location $\vec{r}$ from the perspective of the SSB, at distance $r$ and in direction $\hat{r}$.  It moves with velocity $\bstar = \vec{\beta}_s + \vec{\beta}_p$.  For a star with orbiting companions, $\vec{\beta}_p$ is a complicated function of time, but the magnitude of the perturbations is usually small compared with the space motion of the star.  We can divide the space velocity into the usual proper motion and radial components:

\begin{equation}
\label{betas}
c\vec{\beta}_s = r_0 \vec{\mu} + v_r \hat{r}_0.
\end{equation}

So ignoring the planetary perturbations we have

\begin{equation}
\label{vecr}
\vec{r} = \vec{r}_0 + (r_0 \vec{\mu} + v_r \hat{r}_0) \Delta t,
\end{equation}

\noindent where  $\vec{r}_0$ is the ICRS position of the star at a nominal epoch, and $\hat{\mu}$ points in the direction of proper motion at the epoch of $\vec{r}_0$.  If proper motion is expressed as $(\mu_\alpha,\mu_\delta)$ in units of arc per time (that is, the $\mu_\alpha$ term measures the angular speed in the right ascension direction, not the rate of change in right ascension, which can be large at the poles even for small angles), then $\vec{\mu}$ can be calculated from:

\begin{eqnarray}
\label{coords}
\hat{u} &\equiv& [0,0,1], \\
\hat{\alpha} &=& \frac{\hat{u} \times \hat{r}}{|\hat{u} \times \hat{r}|}, \\
\hat{\delta}& =& \hat{r} \times \hat{\alpha},\\
\vec{\mu} & = & \mu_\alpha \hat{\alpha} + \mu_\delta\hat{\delta},
\end{eqnarray}

\noindent where $\hat{\alpha}$ and $\hat{\delta}$ are constant vectors pointing in the east and north directions for the position $\hat{r}_0$.  We represent the total magnitude of proper motion by $\mu$ (in units of radians yr$^{-1}$, see \citet{wright09}, or the Astronomical Almanac; there are also light-deflection and other corrections that must be made for precise astrometry, but they are not important at the 1 cm/s level \citep{klioner03}).

The correction for (and knowledge of) the systemic velocity of the star, $v_r$, is barely needed, even at 1 cm/s precision.  For decades-long observations of nearby stars, not including it will result in erroneous positions of stars due to its effect on the second-order correction for proper motion (i.e.\ $d \mu/dt$).  The effect on the derived radial velocities due to neglect of this term has magnitude $\beta_\Earth (\Delta t \beta_* / r_0)^2$.  For the extreme case of 30 years of observations of a star 10 light years distant moving at $\beta_* \sim 3\times 10^{-4} \approx 100$ km/s, accumulated positional errors will result in an annual signal with an amplitude of $\approx$ 1 cm/s.

There is also a light travel time effect in play.  Our formulation calculates the true radial velocity to a star as $\vec{\beta}_* \cdot \vec{r_0}$; however this radial velocity is not measured until the light reaches the solar system.  This delay is time variable as the distance to the system changes due to both the space motion of the star and the motion of the star about its system barycenter.  The latter effect requires a careful consideration of the time axis when analyzing measured perturbative radial velocities.  The former is a small effect but can be important for decades-long observations of the nearest, most quickly moving stars at the cm/s level \citep{butkevich14}.  The form of this effect (from Equation D.11 of \citeauthor{butkevich14}) is:

\begin{equation}
\label{lt}
z_{\rm LT} =  v_r \frac{r_0 \mu^2}{c^2} (\Delta t),
\end{equation}

\noindent which is small enough that it can simply be subtracted from $z_B$ if necessary.

\subsubsection{Parallax}

When the wavefronts from this star arrive in the Solar System, they have an approximately spherical locus; the normal to their surface is thus not a constant vector, but depends on the observer's position in the solar system.  Equivalently, the apparent position of the star depends on the observer's position in the solar system, which results in the annual parallax.

We must therefore adapt Eq.~\ref{elementary} to allow for $\hat{r}$ to vary with time, due to both proper motion and parallax.

The Earth orbits the Sun at position $\vec{x}$ from the SSB with velocity \bearth\ (in units of $c$).  The vector from Earth to the star is

\begin{equation}
\label{vecx}
\vec{\rho} = \vec{r}-\vec{x},
\end{equation}

\noindent which has length $\rho$.  The correct form for the Doppler shift is then  (from Eq.~\ref{elementary}):

\begin{equation}
\label{elementaryppm}
\numeas = \nu_{\rm emit}\frac{\gearth (1+z_{\rm GR*})(1+\bearth\cdot\hat{\rho})}{\gstar (1+\zgr) (1+\bstar\cdot\hat{\rho})}.
\end{equation}

The term $(\bearth \cdot \hat{\rho})$ is the velocity of the Earth towards the proper motion and parallax-corrected position of the star, and is the quantity often referred to as the ``barycentric correction'' for coarse measurements.  The term $(\bstar \cdot \hat{\rho})$ is the radial component of the star's motion, and varies with time even in the case of constant \bstar\ (i.e.\ cases where a star has no orbital companions) because of perspective effects due to proper motion and parallax (i.e.\ because $\hat{\rho}$ varies).

\subsubsection{Secular acceleration, the proper-motion-parallax cross term, and second-order planetary perturbations}
\label{crossterms}
We can linearly separate the $(\bstar\cdot\hat{\rho})$ term into three conceptually distinct components:

\begin{equation}
\label{three}
\bstar  \cdot \hat{\rho} =  (\vec{\beta}_s + \vec{\beta}_p) \cdot (\hat{\rho}-\hat{r}) + (\vec{\beta}_s + \vec{\beta}_p) \cdot(\hat{r}-\hat{r}_0) + (\vec{\beta}_s + \vec{\beta}_p)\cdot \hat{r}_0.
\end{equation}

\noindent The final term is the constant, bulk radial velocity $v_r/c$ plus the planetary signal we seek to extract.  The other two terms are ``nuisance'' terms which can be removed.

The middle term in Equation~\ref{three} varies, in part, because $\hat{r}$ changes due to proper motion.  To first order, this causes secular acceleration, a linear increase in redshift as the transverse component of the star's motion gets mixed into the radial component, increasing it with time.  We can see this from the multiplier $(\hat{r}-\hat{r}_0)$, which is simply the angular displacement due to proper motion since the nominal epoch.  So,

 \[
\vec{\beta}_s \cdot (\hat{r}-\hat{r}_0) \approx \vec{\beta}_s \cdot (\vec{\mu} \Delta t) \sim r_0 \mu^2 \Delta t/c.
\]

For a star at 10 light years with $\mu$ = 2\arcsec yr$^{-1}$, the secular acceleration will cause a linear change in the apparent velocity of 0.3 m s$^{-1}$ yr$^{-1}$.  This is the result of transverse motion mixing into the radial direction.  The next order term is the radial motion mixing {\it out} of the radial direction, but this is smaller by a factor of $\mu\Delta t$, resulting in a change of only or 30 micron s$^{-1}$ over 10 years.  So we see that the bulk radial motion of the star can be neglected here.

Contributions from the $\vec{\beta}_p \cdot (\hat{r}-\hat{r}_0)$ term will generally be unimportant.  This term represents the changing perspective on a planetary (or stellar companion) orbit.  For planetary orbits which induce reflex motions of $ |\vec{\beta}_p | < 10^{-6} $, this will result in changes of order $\mu \Delta t |\vec{\beta}_p| < 10^{-10}$, or $< 1$ cm/s, and so can be neglected here.  For stars in binary systems with larger orbital motions, this term could be included as part of a complete model of the stellar system.

The multiplier (the second factor) in the first term of Equation~\ref{three} represents the parallax, $\varpi$, of the star (that is, its angular displacement from its SSB position due to the physical displacement of the Earth), and so has a magnitude of $\varpi < 10^{-5}$ (radians) for all real stars.  For 1 cm/s precision, as with the proper motion term, it is sufficient for all real stars to approximate this quantity as being equal in magnitude and opposite to the projection of the Earth's displacement on the sky:

\[
\hat{\rho}-\hat{r} \approx -\left(\frac{\vec{x}}{r} - \frac{ \vec{x}\cdot\hat{r}}{r}\hat{r}\right).
\]

Since (in this approximation) this vector has no radial component, the product will be non-negligible only in plane-of-sky directions:

\[
\vec{\beta}_s \cdot (\hat{\rho}-\hat{r})\approx -\frac{r_0 \vec{\mu}}{c} \cdot \frac{\vec{x}}{r} \approx -\frac{\vec{\mu}}{c} \cdot \vec{x}.
\]

This is the ``parallax-proper motion'' cross term.  Its origin is the transverse motion of the star which is mixed into the radial direction by the varying perspective generated by the Earth's motion about the Sun (that is, the variation of the $\hat{\rho}$ vector due to variations in $\vec{x}$) .  It inherently couples the star's intrinsic motion to the Earth's motion, blurring what might otherwise be a clean separation between corrections for the Earth's motion and the star's motion.  This term is typically very small, being of order $|\beta_s \varpi|$, so as large as 10 cm/s for the nearest ($\varpi \sim 10^{-5}$) disk stars ($|\beta_s| \sim 10^{-4}$).  As with the proper motion term, the corresponding radial component is smaller by another factor of $\varpi$, and so is clearly negligible.

Even more than with the proper motion term, the $\vec{\beta}_p \cdot (\hat{\rho}-\hat{r})$ can be neglected for all real single stars, but may be important at the 1 cm/s level for some nearby binary systems.

So, even for work at 1 cm/s, both astrometric and radial velocity ``wobbles'' due to planetary companions can be ignored in the barycentric correction process (that is, $\vec{\beta}_p$ does not produce important cross terms with $\vec{x}$ or $\vec{\mu}$).  Exceptions would be any system for which the orbital motions produce velocities for which $\beta_p \varpi$ is above the desired precision.  For instance, in a binary system at 20 pc ($\varpi \sim 10^{-6}$) with orbital velocity 30 km/s ($\beta_p \sim 10^{-4}$), this term would have magnitude $10^{-10}\sim 3$ cm/s.

\subsection{General formulation for 1 cm/s precision}

Precise Doppler measurements of radial velocities compare a spectral feature's measured frequency, $\numeas$, to some reference frequency, $\nu^\prime$.  Each measurement thus reports

\begin{eqnarray}
\label{zmeaseq}
\zmeas & \equiv & \frac{\lmeas-\lambda^\prime}{\lambda^\prime}\\
&=& \frac{\nu^\prime-\numeas}{\numeas}.
\end{eqnarray}

From. Eqs.~\ref{nuprime}~\&~\ref{elementaryppm} we have:

\begin{equation}
  \zmeas = \frac{(1+\bstar\cdot\hat{\rho}) (1+\zgr)}{(1+\bearth\cdot\hat{\rho})\gearth(1+z^\prime)}-1.
\end{equation}

We wish to transform this to $z_{\rm true}$, and to eliminate the $z^\prime$ term. Examination of Eq.~\ref{ztrue} allows us to do this:

\begin{equation}
\label{nohelp}
\frac{(1+z_{\rm true})}{(1+\zmeas)} =  \left(\gearth\frac{(1+\bearth\cdot\hat{\rho})}{ (1+\zgr)}\right)\left(\frac{(1+\bstar\cdot\hat{r}_0) }{(1+\bstar\cdot\hat{\rho})}\right) = (1+z_B).
\end{equation}

The multiplicand (first factor) in the middle of this equation represents the correction to a measured redshift one must apply to remove the (time variable) effects of the moving platform the Earth affords the observatory.

As written, the second fraction in middle of Eq.~\ref{nohelp} is unhelpful:  it converts the radial motion of the star (due to both space motion and planetary motion) from the vantage point of the Earth to what would be observed along a constant vector.  But this term cannot be calculated precisely because \bstar\ includes $\vec{\beta}_p$, which is the very thing we wish to deduce from the observations.

But, as we have argued in Section~\ref{crossterms}, the effects of a shift in perspective on the measured planetary perturbations are generally too small to measure, even at 1 cm/s precision, so we are justified using this term to correct only for the (small) parallax-proper motion and secular acceleration terms (that is, we can substitute $\bstar \rightarrow \vec{\beta}_s$).

This gives us our desired equation for $z_B$:

\[
z_B \approx \left(\gearth\frac{(1+\bearth\cdot\hat{\rho})}{ (1+\zgr)}\right)\left(\frac{(1+\vec{\beta}_s\cdot\hat{r}_0) }{(1+\vec{\beta}_s\cdot\hat{\rho})}\right) - 1.
\]

That is, to 1 cm/s precision, the redshifts one would measure for a set of spectral features with respect to some fiducial set of frequencies $\nu^\prime$ from an optimal vantage point can be derived from the redshifts we actually measure with respect to that same $\nu^\prime$ by using $z_B$ in the equation above and Equation~\ref{zb}.

There are two corrections to consider that are not in this calculation: the effects of the Shapiro delay (Section~\ref{sec:elementary}) and light travel time (Section~\ref{lighttravel}).  The Shapiro delay can be written

\begin{equation}
v_{\rm SD} = \frac{d}{dt} \left(-\frac{2GM_\odot}{c} \ln {(1-\hat{r}_\odot \cdot\hat{\rho})}\right),
\end{equation}

\noindent where $\vec{r}_\odot$ is the vector from Earth to the Sun. In our code we implement this approximate form:

\begin{equation}
z_{\rm SD} = -\frac{2 G M_\odot}{c^2} \frac{\bearth \cdot \left[\hat{\rho} - (\hat{x}\cdot\hat{\rho})\hat{x}  \right]}{ \lvert \vec{x} \rvert (1+\hat{x}\cdot\hat{\rho})} \label{extra}.
\end{equation}

\noindent  This term, along with the light travel term $z_{\rm LT}$ from Equation~\ref{lt}, is small enough that it can be subtracted from $z_B$ for additional precision.  For completeness (and maximum agreement with TEMPO2) our code includes these small corrections, reporting

\begin{equation}
\label{BC}
z_B \approx \left(\gearth\frac{(1+\bearth\cdot\hat{\rho})}{ (1+\zgr)}\right)\left(\frac{(1+\vec{\beta}_s\cdot\hat{r}_0) }{(1+\vec{\beta}_s\cdot\hat{\rho})}\right) - 1 - z_{\rm LT} - z_{\rm SD}.
\end{equation}
\section{BARYCORR}
\label{sec:barycorr}

We have written a publicly-available code that implements the correction described in this paper. \verb#ZBARYCORR# uses the JPL DE405 ephemeris\footnote{In principle, closed form approximations for the Earth's position, such as the IAU Standards of Fundamental Astronomy (SOFA) {\tt EPV00} \citep{wallace04} are sufficient for cm/s precision, but would dominate our error.} to calculate the positions and velocities of all the relevant solar system objects, applies the rotation, precession, nutation, and polar motion of the Earth -- using the Earth orientation parameters supplied by the International Earth Rotation and Reference System (IERS) and interpolated with Craig Markwardt's \verb#EOPDATA#\footnote{\url{http://cow.physics.wisc.edu/~craigm/idl/idl.html}} -- applies the stellar motion, Shapiro delay, and light travel term, and finally calculates the quantity $z_B$ --- the barycentric correction independent of the measured redshift. Since the true radial velocity depends on \zmeas\ (see Eq. \ref{zb}), we also include \verb#BARYCORR#, a wrapper for \verb#ZBARYCORR# that requires the measured redshift and returns the corrected velocities.

These codes build off the California Planet Search barycentric correction code (see acknowledgements), the time utilities from \citet{eastman10b}, as well as routines written by Craig Markwardt\footnote{\url{http://cow.physics.wisc.edu/~craigm/idl/ephem.html}}. They are now included as part of the EXOFAST package \citep{eastman13}\footnote{\url{http://astroutils.astronomy.ohio-state.edu/exofast/}} and can also be found at exoplanets.org\footnote{\url{http://exoplanets.org/code/}}, along with the RVLIN \citep{wright09} and BOOTTRAN \citep{wang12} orbit-fitting packages.
\subsection{Comparison with TEMPO2}

We have calculated barycentric corrections for the nominal case of $\tau$ Ceti observed at CTIO (see Section~\ref{errors} for details) with our code that performs the calculations described throughout this paper. To validate our procedure and code, we also performed the same correction with TEMPO2 (see Appendix \ref{sec:tempo}). Figure~\ref{TEMPO} shows that the difference between the two codes has an RMS of 0.037 cm/s and a peak to peak difference of 0.24 cm/s --- much better than 1 cm/s over the 14 year span with no long term trend (but see below). The remaining residuals are strongly correlated with the sidereal day, as shown in Figure \ref{fig:siderealday}, suggesting that the dominant remaining effect is likely due to an undiagnosed shortcoming in our treatment of the Earth's orientation. While these residuals also phase well with the tropical year, other comparisons with TEMPO2 and our code (not shown) have allowed us to determine this is likely an alias with the residuals at the sidereal day.

\begin{figure}
\includegraphics{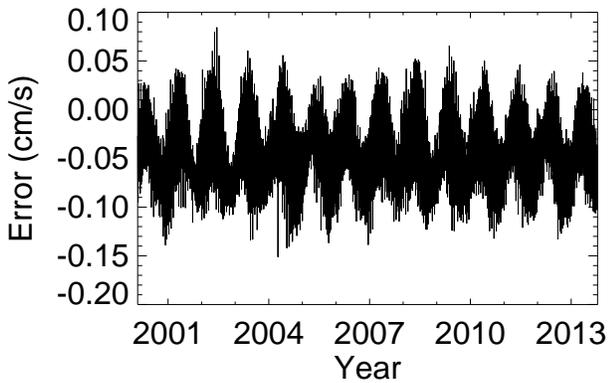}
\caption{Comparison of the barycentric correction generated by {\tt TEMPO} and generated with our code using the method described in this paper over a 14 year period. The residuals have an RMS of 0.037 cm/s and a peak to peak difference of 0.24 cm/s with no long term trend. The residuals phase with the sidereal period, (see Figure \ref{fig:siderealday}), suggesting that the dominant remaining effect is in our treatment of the Earth's orientation.}
\label{TEMPO}
\end{figure}

\begin{figure}
\includegraphics{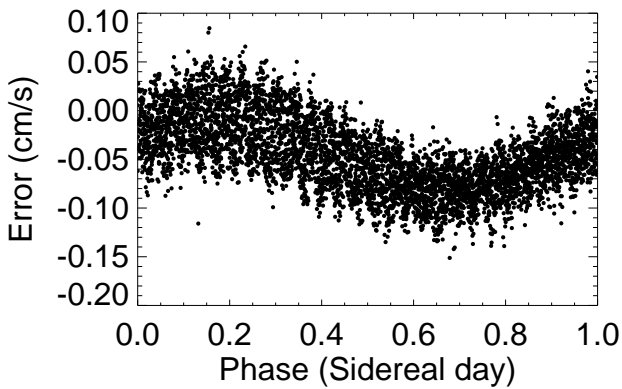}
\caption{The residuals from Figure \ref{TEMPO} phased to the sidereal day. The strong coherence and sinusoidal nature suggests that the remaining difference is dominated by our treatment of the Earth's orientation.}
\label{fig:siderealday}
\end{figure}

An important caveat to the use of TEMPO2's prediction mode to validate barycentric correction algorithms or perform barycentric corrections is that TEMPO2 in prediction mode requires the user to specify the {\it observed} frequency and frequency drift (i.e., the observed pulsar period and its time derivative, $\dot{P}$, in the parameters \verb#F0# and \verb#F1#) at a given epoch (see Appendix). TEMPO2 can thus not be used in this way to validate one's calculation of long-term linear terms.

\begin{figure}
\includegraphics{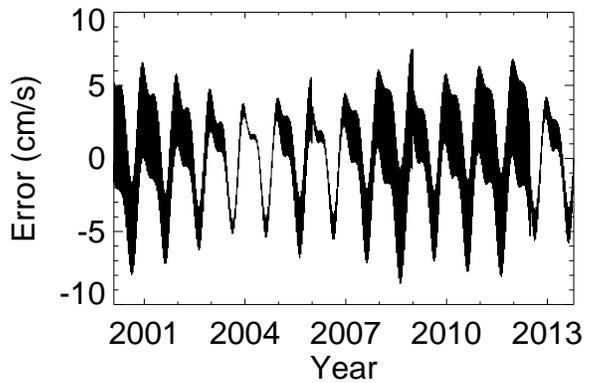}
\caption{Comparison of the barycentric correction generated by {\tt TEMPO2} and generated with the previous version of the barycentric velocity correction code used by the California Planet Survey. A secular trend and zero point offset have been subtracted, because these terms are not included in the old code (they were subtracted by later parts of the Doppler analysis pipeline) and because they cannot be validated using TEMPO2 (see \S 5.1)}
\label{fig:oldcode}
\end{figure}

Figure \ref{fig:oldcode} compares TEMPO2 and the previous version of the barycentric velocity correction code used by the California Planet Survey team. It shows an RMS residual of 3 cm/s and a peak to peak residual of over 16 cm/s. The primary term missing from the old barycentric correction code is the proper-motion parallax cross term. While the contribution of the barycentric correction algorithm to the error budget of velocities published by the California Planet Survey is over an order of magnitude better than its historical measurement uncertainties, it is insufficient for the next generation of ultra-precise radial velocity work. Our new code is nearly two orders of magnitude better than the previous version.

Table~\ref{table} contains an example of the barycentric corrections both TEMPO2 and our code calculate for $\tau$ Ceti from CTIO shown in figures \ref{TEMPO} and \ref{fig:siderealday}. The full table of 5432 values spanning roughly 14 years is distributed with our code for a more detailed comparison.

\begin{deluxetable}{rrr}
\tablecolumns{3}
\tablecaption{Barycentric corrections, $z_B$, for $\tau$ Ceti from CTIO calculated with TEMPO and our code for the first 5 times in our example time series. The full table of 5432 values can be found in the online table for a more detailed comparison. Since the presumed measured redshift, \zmeas, is zero, these are converted to velocities by multiplying by $c$ (Eq. \ref{zb}) when the differences are plotted in figures \ref{TEMPO} and \ref{fig:siderealday}.
\label{table}
}
\tablehead{\colhead{Date} & \colhead{TEMPO} & \colhead{This algorithm}\\
\colhead{JD$_{\rm UTC} - 2400000$} & \colhead{} & \colhead{}}
\startdata
51581.00000000000000 & -0.00007942787937 & -0.00007942788026 \\
51581.92064076615497 & -0.00007925377190 & -0.00007925377301 \\
51582.84128153184429 & -0.00007911673755 & -0.00007911673949 \\
51583.76192229799926 & -0.00007887624637 & -0.00007887624877 \\
51584.68256306368858 & -0.00007841065462 & -0.00007841065780 \\
\enddata
\end{deluxetable}

\section{Magnitude of Terms}
\label{sec:magnitude}
\subsection{Importance of including various analytic terms}

At the 10 m/s level, a simple, Galilean barycentric correction (the $(\bearth\cdot\hat{r})$ term in Eq.~\ref{BC}) is a sufficiently precise formulation of the problem, although for a few stars over long time spans the secular acceleration term may become important.

At the 1 m/s level, one must account for the multiplicative form of Eq.~\ref{zb} (the cross term $z_B\zmeas$).  This term has typical magnitude $c\bearth^2 \sim$ 3 m/s for most stars, since $\zmeas$ is dominated by the barycentric correction for most targets.  This term is not, strictly speaking, a ``relativistic term'' since it can appear in a classical formulation of the Doppler problem that invokes the luminiferous ether; it is best thought of as the second-order correction to the simplest form of the Doppler formula, $\Delta \nu/\nu = v/c$.

For many nearby stars with significant proper motion, the final, secular acceleration term is also important.  Both the time dilation and General Relativistic (GR) terms are of sufficient magnitude to include here, but are constant at the 1 m/s level, so are not important.

At the 10 cm/s level the parallax-proper-motion term, explained in Section~\ref{crossterms}, becomes important for nearby stars, but can be neglected for distant ones.

At 1 cm/s, all of the remaining terms are important.  The GR redshift at the position of the Earth is $\approx 3$ m/s, and is modulated by the eccentricity of the Earth's orbit ($\sim 0.01$), so is an important term to include at the $\sim 3$ cm/s level.  Of similar magnitude and sign is the variation in the time dilation term of the Earth, which is of order $ce\bearth^2 \sim 3$ cm/s (the fact that these terms are similar is a consequence of the virial theorem).

Below 1 cm/s, the Shapiro delay and light travel time effects should be included; the approximations and assumptions that went into this calculation should be revisited (such as the approximation that $\vec{\beta}_* \rightarrow \vec{\beta}_s$); and a fully general-relativistic model including both the Earth and star's motion should be used.
\subsection{Uncertainty incurred from errors in various quantities}

\label{errors}

In the figures in this section, we compare our barycentric code's output with its output with certain effects subtracted or altered.  $\tau$ Ceti is the nominal target, and the nominal observatory is the CTIO 1.5m.  The location of $\tau$ Ceti near both the Celestial Equator and the ecliptic makes it a nearly ``worst-case scenario'' for errors in barycentric correction.  To ensure that these figures efficiently probe and illustrate diurnal, annual, and super-annual effects (e.g.\ nutation), most figures show one point every 22.09 hours over a 14-year period from 2000 to 2014. For errors that are only relevant on year timescales, such as timing or coordinate errors, we only show a 500 day range. And for errors in the observatory position, which repeat every sidereal day, we only show one day. Note that some of these times occur during the day; the figures show them as a test of the algorithm, not as examples of real barycentric corrections.

\subsubsection{Earth Rotation and Orientation}

\label{EOP}
To achieve 1 cm/s precision in barycentric correction, it is necessary to account for the precession, nutation, and variable rotation of the Earth.  The first two of these, representing the motion of the rotation axis of the Earth in response to torques from lunar and solar tides, can be predicted in advance with sufficient fidelity for 1 cm/s barycentric correction precision.

The changes in the angular speed of the Earth are not perfectly predictable, and must be accounted for retrospectively.  These variations are encoded in the difference between the time standards UT1 and UTC, and in the leap seconds that are periodically added to civil time.  These values are carefully tracked by the International Earth Rotation Service (IERS), which provides regular bulletins on their values.  These bulletins also give precise, retrospective values for the precession, nutation, and polar motion\footnote{The polar motions are rotations of the body axis of the Earth with respect to its rotation axis, i.e.\ the drift of the location where the rotation axis intersects the Earth's surface.  The dominant, quasi-periodic modes of polar motion are called the Chandler Wobble.  These, and the unpredictability of leap seconds, are due to changes in the Earth's moment of inertia due to both internal motions and the redistribution and motion of material on the Earth's surface.} , but these higher-order corrections are not necessary for 1 cm/s precision. By dividing our desired 1 cm/s precision by the Earth's rotational speed (460 m/s), we see that errors up to 4.4\arcsec\ are acceptable. Note that the primary effect of the variable angular speed of the Earth on the barycentric correction is not that the speed of the observatory is variable, but that the accumulated phase error in the rotation of the earth will yield an error in the direction of its motion.

To account for all three of these motions, our code employs the routines \verb#HPRSTATN# and \verb#HPRNUTANG# written by Craig Markwardt\footnote{\url{http://cow.physics.wisc.edu/\%7Ecraigm/idl/idl.html}} which use the routine \verb#EOPDATA# to parse IERS Bulletin A\footnote{The bulletin at {\tt ftp://maia.usno.navy.mil/ser7/finals.data} must be saved as {\tt IERS\_final\_a.dat} in the directory defined by the environment variable {\tt ASTRO\_DATA}.}, which includes the UT1$-$UTC corrections.  Optionally, \verb#HPRNUTANG# can also use the JPL ephemeris to determine the retrospective nutation angles.

\begin{figure}
\includegraphics{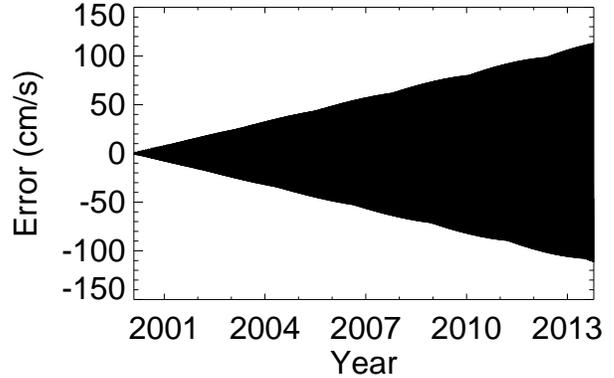}
\caption{Error incurred by ignoring the effects of precession in a barycentric motion calculation.  Precession cannot be ignored, even for work at 1 m/s.  The point density in this plot makes the diurnal variation appear as a solid black region.}
\label{precession}
\end{figure}

\begin{figure}
\includegraphics{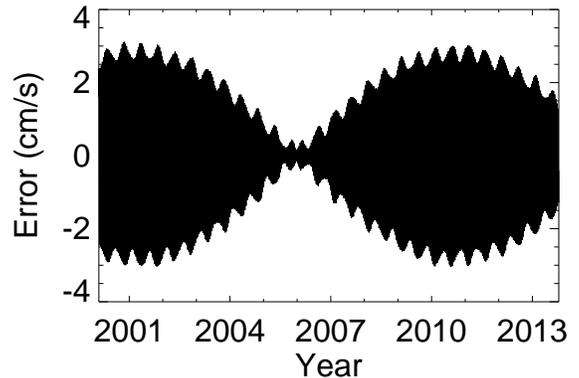}
\caption{Error incurred by ignoring the effects of nutation in a barycentric motion calculation.  Nutation should not be ignored for work at 10 cm/s.  The point density in this plot makes the diurnal variation appear as a solid black region}
\label{nutation}
\end{figure}

\begin{figure}
\includegraphics{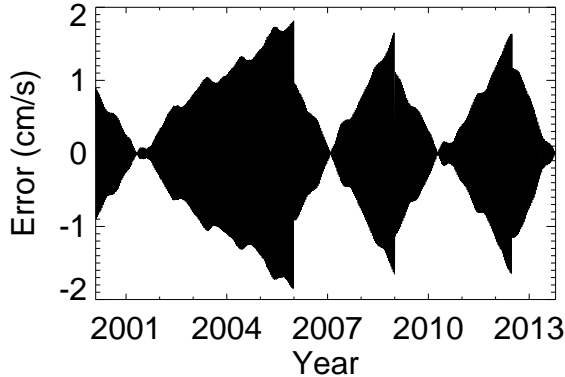}
\caption{Error incurred by ignoring the UT1$-$UTC corrections in a barycentric motion calculation.  To be conservative, this term should not be ignored for work near 10 cm/s, although it is a factor of a few below that limit.  The point density in this plot makes the diurnal variation appear as a solid black region.  The effects of leap seconds can be seen in the discontinuous nature of the magnitude of the diurnal variations. }
\label{UT1}
\end{figure}

Figure~\ref{precession} shows the effects of neglecting the precession, and Figure~\ref{nutation} shows the effects of neglecting the nutation of the Earth's rotation axis.  Figure~\ref{UT1} shows the effects of neglecting the variations in the Earth's angular speed (i.e.\ the UT1$-$UTC correction).   Plotted is the difference in the measured barycentric correction towards $\tau$ Ceti from CTIO between our code using the full Earth Orientation Parameter (EOP) description, and with our code using the \verb#no_precession# and \verb#no_nutation# options in \verb#HPRNUTANG#.  These figures show that precession is important even at the 1 m/s level, nutation is important at the 3 cm/s level, and the UT1 correction, in practice, is important at 2 cm/s.\footnote{In principle, the maximum UT1$-$UTC offset of 0.9 seconds would imply a $\sim3$ cm/s error, and if current proposals to abolish leap seconds are successful this error would grow indefinitely.} No work claiming detections with amplitudes below 10 cm/s should neglect these effects.

\begin{figure}
\includegraphics{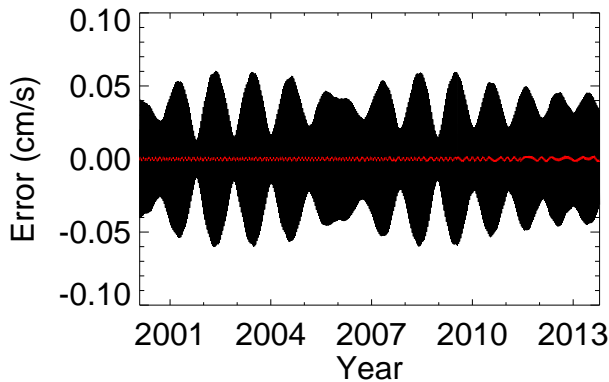}
\caption{Error incurred by ignoring the effects of polar motion (black) and of using predictive formulae for the precession and nutation (red) in a barycentric motion calculation.  These terms are not important, even for work at 1 cm/s, and the error introduced from using the predictive formula is clearly negligible. The point density in this plot makes the diurnal variation appear as a solid black region.}
\label{polar}
\end{figure}

Higher order effects, for instance from polar motion, are not relevant to radial velocity work at 1 cm/s precision.  Figure~\ref{polar} shows the effects of neglecting polar motion and of using formulaic (i.e.\ predicted) precession and nutation instead of retrospectively measured values.  All three approaches are valid at the 1 cm/s level, with the polar motion being the next most important term.

\subsubsection{Time of Observation}

There are two important considerations when reducing the time of an observation recorded at the observatory to other rigorously defined measures of time.  The first is that due to light travel time effects within the Solar System, one should analyze reduced radial velocity measurements according to the times that the signals measured arrived at the SSB --- i.e., the Barycentric Julian Date in Barycentric Dynamical Time (\bjdtdb). Such calculations are discussed in \citet{eastman10b}.

The second is that one must rigorously determine the motion of the observatory towards the target at the moment and place of observation, requiring reduction of observatory time to Barycentric Dynamical Time for use in an ephemeris of the Earth's motion, and proper conversion to UT1 for use in determining the Earth's orientation.  \citet{eastman10b} provides a good discussion of these reductions.

\begin{figure}
\includegraphics{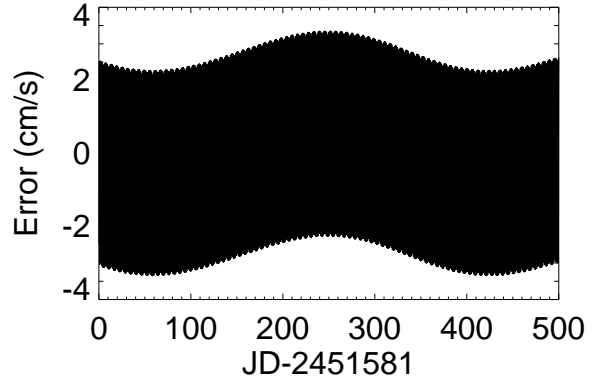}
\caption{Error incurred from a timing error of 1 s on the barycentric correction. Note that we have only shown 500 days, as the effect repeats every year. For 1 cm/s barycentric corrections, the effective time of an observation must be known to 250 ms.  Achromatic seeing may make the effective time of an observation wavelength-dependent, posing a technical challenge for future radial velocity work at 10 cm/s.  The point density in this plot makes the diurnal variation appear as a solid black region.}
\label{timeerr}
\end{figure}

At 1 cm/s precision, the time of observation must be known to a fraction of a second for good barycentric correction.  The error incurred comes primarily from the rotation of the Earth, and this causes errors as large as $v_{\rm rot} 2\pi/{\rm day} = 3.4$ cm/s/s.  Figure~\ref{timeerr} shows the errors incurred from miscalculating the time of observation of $\tau$ Ceti at CTIO by 1 second.

This has two implications:  the first is that the reduction of observatory time to terrestrial dynamical time and (more importantly) UT1 must be performed to a precision and accuracy of better than 250 ms to ensure 1 cm/s barycentric precision, so both leap seconds and the (UTC-UT1) terms must be properly accounted for.  This is equivalent to our discussion in Section~\ref{EOP}.

The second, more difficult-to-implement implication, is that the effective time of a potentially lengthy observation, often tens of minutes long, must be determined to a quarter of a second.  This is normally accomplished by means of an exposure meter, which monitors the flux of starlight through the spectrograph (i.e.\ the exposure meter sits {\it behind} the slit).  The temporal photocenter (i.e.\ time of arrival time of the median photon) of the observation is then a good approximation for the effective exposure time center, although at 1 cm/s precision higher-order corrections may be necessary, especially under poor observing conditions.  Simple geometric exposure time midpoints are likely unreliable at 1 cm/s precision since seeing and guiding variations can make the temporal photocenter and midpoint time differ by more than 250 ms.  {\bf Note that atmospheric chromatic aberration can make time-variable slit losses (or, equivalently, fiber coupling losses) wavelength dependent, which may require monitoring exposure photocenters as a function of wavelength.}

\subsubsection{Stellar coordinates and parallax}

\begin{figure}
\includegraphics{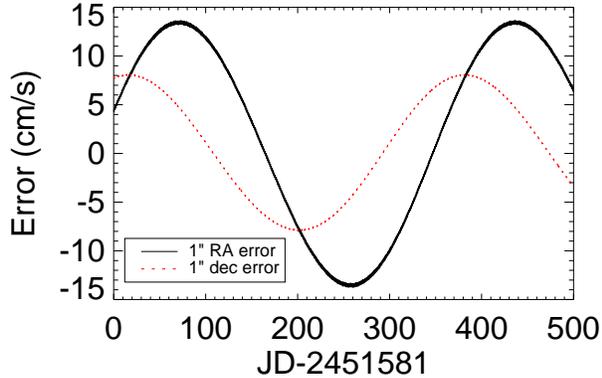}
\caption{Error incurred from a 1\arcsec\ error in stellar position for $\tau$ Ceti. Note that we have only shown 500 days, as the effect repeats every year. Stellar positions must be better than 1\arcsec\ for work at or below 10 cm/s.}
\label{Star1as}
\end{figure}

Figure~\ref{Star1as} shows the effects of errors in stellar coordinates.  Here, we have applied 1\arcsec\ of error in right ascension and declination to $\tau$ Ceti.  We see that Stellar positions must be much better than 1\arcsec\ for work at or below 10 cm/s. Indeed, we can simply divide the required precision (1 cm/s) by the Earth's orbital speed (30 km/s) to get the angular precision necessary in the stellar coordinates (69 mas). In particular, stars with large but poorly-known proper motions may accumulate right ascension uncertainty large enough to affect work at 10 cm/s.

\begin{figure}
\includegraphics{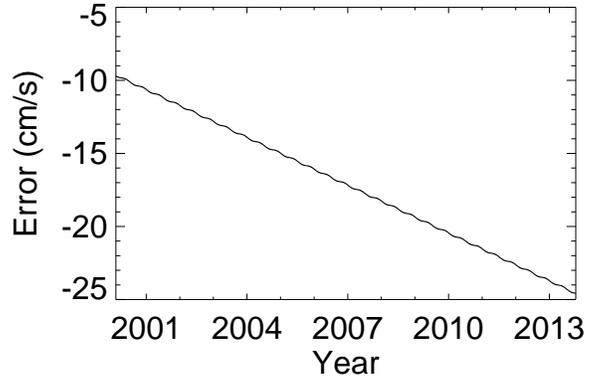}
\caption{Error incurred from a 10 mas error in parallax of $\tau$ Ceti (which is 274 mas).  The primary effect is to introduce an anomalous secular acceleration, which might be mistaken for a long-period companion.}
\label{par}
\end{figure}

Figure~\ref{par} shows the effects of an error in the parallax.  The primary effect of error in parallax is that the barycentric correction miscalculates the space motion of the star from the proper motion, and thus under- or over-predicts the mixing of the tangential component into the radial direction via the stellar proper motion.  The result is an anomalous secular acceleration of the stellar motion.  Such a signal could be mistaken for a long-period orbital companion.

\subsubsection{Stellar velocity and proper motion}

\begin{figure}
\includegraphics{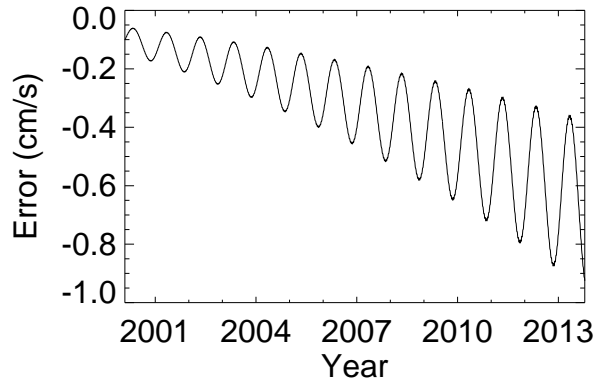}
\caption{Error incurred from an error in the systemic velocity of 100 km/s of $\tau$ Ceti. For 1 cm/s barycentric corrections over a decade we can safely ignore systemic velocities up to 100 km/s.}
\label{fig:rv}
\end{figure}

The primary effect of the radial component of the space motion of the star is to produce a constant redshift.  The next most important effect is in the mixing of this motion out of the radial direction via the proper motion of the star, and this, as Section~\ref{crossterms} showed, is negligible even at 1 cm/s. We demonstrate this in Figure \ref{fig:rv} by showing the difference between no radial velocity (our nominal example for $\tau$ Ceti) and $\tau$ Ceti with a fictional 100 km/s systemic velocity.

\begin{figure}
\includegraphics{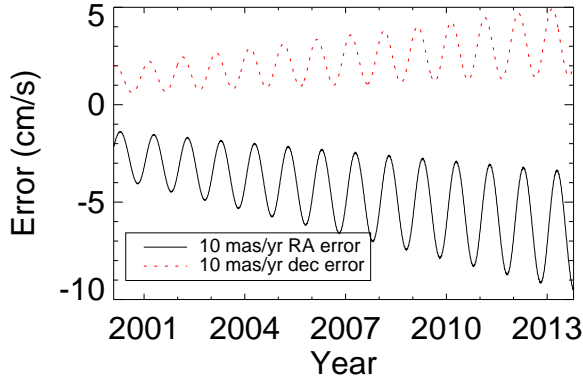}
\caption{Error incurred from a 10 mas/yr error in the proper motion of $\tau$ Ceti. Parallax must be known to better than 10 mas/yr for work at 10 cm/s.}
\label{pm}
\end{figure}

Errors introduced by errors in proper motion can be large, because they can accumulate over time to large errors in position.  Figure~\ref{pm} shows the effect of a 10 mas/yr error in the proper motion of $\tau$ Ceti.  The primary effects of errors in proper motion are in the magnitude of the parallax-proper motion term as this tangential motion mixes into the radial motion as the star moves from both parallax and proper motion; and in the coordinates of the star.  These three effects can be of similar magnitude; Figure~\ref{pm} shows that they all matter at around the 5 cm/s over decadal timescales.

\subsubsection{Telescope Coordinates and Earth Figure}

The telescope position with respect to the Earth center must be well known for a proper barycentric correction to be made. The equatorial rotational velocity of the Earth is 460 m/s, so an error of 100 m for an equator-based observatory corresponds to an error in time of 0.2 s, just at the tolerance required for 1 cm/s precision.  This is a rather generous requirement, since most astronomical observatories have positions known with significantly better fidelity than this.  \citet{mamajek12} provides a good description of the various methods and coordinate systems for determining observatory positions, applied to the specific problem of CTIO.

\begin{figure}[!htbp]
\includegraphics{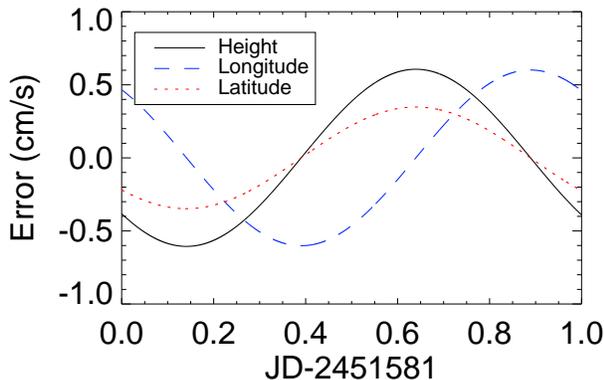}
\caption{Error incurred from a positional error of 100 m on the barycentric correction for the CTIO 1.5-m. We only show one day because the effect repeats daily with no long-term trend. For 1 cm/s barycentric corrections, 100 m is sufficient positional accuracy.}
\label{Coordinates}
\end{figure}

Figure~\ref{Coordinates} shows the errors associated with having the coordinates of the CTIO 1.5-m telescope incorrect by 100 m in longitude, latitude, and in height.  All 3 cases produce similar errors, all below 1 cm/s.

The Earth is not perfectly spherical, but is well represented by an oblate ellipsoid.  One must account for this oblateness of the Earth to avoid errors of several km in position.  Higher order corrections (the higher-order spherical harmonic amplitudes that describe the geoid) are important at 100 m, and so are just barely required for 1 cm/s barycentric correction precision.  Radial velocity work at 1 cm/s would have to take these into account, but work at 10 cm/s may safely ignore them.
\acknowledgements

We thank Debra Fischer for sponsoring this project, and a careful read of the manuscript.  We gratefully acknowledge NSF grant AST1109727 and NASA grant NNX12AC01G (PI Debra Fischer).

The original barycentric correction procedures (``the bary code'') were written in IDL by many members of the California Planet Survey, primarily Chris McCarthy, Geoff Marcy, and Jeff Valenti, with preliminary work by Mario Savio.  Important conceptual contributions, including the importance of relativistic terms, were made by John Johnson, Eric Ford, and Debra Fischer.

This paper was informed by valuable discussions with the original authors of the bary code, especially Chris McCarthy.  We have also had useful conversations with Nikita Roytman and Sabine Reffert on the subject of barycentric corrections.

J.T.W.\ thanks Alex Wolszczan for his help in using TEMPO2  and Paul Demorest for hosting him in Charlottesville and allowing him to pester him in person over the course of two days (and over email for a much longer period) in our quest to understand and implement TEMPO2.  J.T.W.\ thanks Andrea Lommen and especially Nathaniel Garver-Daniels and for their assistance installing TEMPO2 on a Mac.  We thank Eric Mamajek for his inordinate efforts in ascertaining the proper and precise coordinates of the CTIO 1.5m telescope.

The Center for Exoplanets and Habitable Worlds is supported by the Pennsylvania State University, the Eberly College of Science, and the Pennsylvania Space Grant Consortium.

This research has made use of NASA's Astrophysics Data System.

\appendix
\section{TEMPO2}
\label{sec:tempo}

The pulsar community has used the TEMPO2 code \citep{hobbs06,edwards06} to great success to time pulsar pulse arrivals for decades, to a precision far greater than required even for 1 cm/s observations.  Indeed, the discovery of the first exoplanets by \citet{wolszczan92} was made with this package using a technique analogous to Doppler planet detection, and these planets are smaller than any other exoplanets discovered to date.

Today, the state-of-the-art package is TEMPO2\footnote{Currently maintained at \url{http://sourceforge.net/projects/tempo2/}, with documentation at \url{http://www.atnf.csiro.au/research/pulsar/tempo2/}}, which incorporates all geometric and special relativistic terms in the solar system and the target system, including the effects of orbital companions.  TEMPO2 thus provides a test for other barycentric correction codes, allowing one to determine their utility and precision.

\subsection{Setting up parameter files}

TEMPO2 derives its inputs from a text file named for the star.  Since TEMPO2 expects to work on pulsar data, and since pulsars are named by their coordinates, this file must have a digit as its first character.  An example of our parameter file for $\tau$ Ceti, \verb#10700.par#, is given below:

\begin{verbatim}
PSR                   10700
RA           +01:44:05.1275
DEC          -15:56:22.4006
PMRA               -1721.05
PMDEC                854.16
PX                   273.96
PMRV                0.00000
F0         1.00000001550505
F1         -3.273402529e-17
PEPOCH          48348.56250
POSEPOCH        48348.56250
DMEPOCH         48348.56250
DM                        0
EPHEM                 DE405
CLK                  UNCORR
TZRMJD          48348.56250
TZRFRQ            545000000
TZRSITE                 coe
\end{verbatim}

In these files, the RA and Dec are given in the epoch defined by POSEPOCH in hours and degrees, respectively.  Proper motions are given in units of milliarcseconds per year, and the parallax in milliarcseconds.  The frequencies F0 and TZRFRQ are given in Hz, and the time derivative of F0, F1, is given in Hz s$^{-1}$.  The three epochs correspond to that of the pulsar frequency, the position of the star, and the dispersion measure (\verb#DM#, which is set to zero here because it is not relevant to barycentric corrections of stars).  They are given as MJD, and in this file represent the definition of the Hipparcos epoch (approximately J1991.25).  The TZ terms refer to the observatory location, epoch, and frequency of a reference pulse arrival time (this provides phase information of the pulse arrivals; since we are only concerned with the frequency of the pulse arrivals, the values of these parameters are not relevant to barycentric correction calculations).

In our nomenclature, the TEMPO2 arrival frequency \verb#F0# is the frequency of light as it enters the Solar System at a given epoch, in the frame of the SSB in hertz (and not $\nu_{emit}$, which corresponds to the frequency of the "pulsar" in its own frame).  We will choose this to be our fiducial frequency $\nu^\prime$.  The value 1.00000001550505 Hz corrects for the difference between the input time in Barycentric Dynamical Time (TDB) and the time in Barycentric Coordinate Time (TCB).  The parameter \verb#F1# nominally represents the pulsar spin-down time, which is zero in our idealized case, but it also contains the secular acceleration \citep[which pulsar astronomers know as the Shklovski\u{\i} effect;][]{shklovskii70}.  We therefore set it to $-$\verb#F0#$r_0\mu^2/c$.

One need not specify $v_r$ because, as we have seen, it does not enter the problem.  It can be included, however, using the \verb#PMRV# parameter (measured as $v_r/r_0$).  We may further choose to reference all frequencies to $\nu(t_0)$, so this specifies $\nu^\prime = \nu(t_0)$ and $z^\prime=0$.

The parameter \verb#EPHEM# refers to the name of the JPL ephemeris to be used, here set to DE405.  The CLK parameter refers to the clock corrections to be used to interpret clock times at a radio observatory; it is not relevant to barycentric correction work and so is set to UNCORR, which instructs TEMPO2 to ignore these corrections.

The coordinates of the telescope should go into a user-created file in the \verb#observatory/# directory.  Our file optical.dat contains:

\begin{verbatim}
# New format observatory information file.
#
1814985.3 -5213916.8 -3187738.1 CTIO_1.5-meter ctio
\end{verbatim}

\noindent This position corresponds to the position determined by \citet{mamajek12}.

For binary systems such as $\alpha$ Centauri, for which coordinates vary by more than a simple proper motion correction, we generate this file dynamically for every call to TEMPO2.

\subsection{Maintaining EOP and other files}

Users of TEMPO2 must maintain files that track variations in the Earth's rotation as they are measured.  These files can be accessed using \verb#git# and \verb#CVS# through the TEMPO2 project portal.\footnote{At \url{http://tempo2.sourceforge.net}}

The files that must be maintained for TEMPO2 to properly function as a barycentric correction calculator are the \verb#ut1.dat# file maintaining the TAI$-$UT1 difference\footnote{\url{http://sourceforge.net/p/tempo/tempo/ci/master/tree/clock/ut1.dat} and goes in the local {\tt clock/} directory}; the \verb#leap.sec# file \footnote{\url{http://sourceforge.net/p/tempo/tempo/ci/master/tree/clock/leap.sec} and also goes in the local {\tt clock/} directory} which maintains a list of leap seconds for \verb#-tempo1# compatability mode, and the \verb#utc2tai.clk file#\footnote{\url{http://tempo2.cvs.sourceforge.net/viewvc/tempo2/tempo2/T2runtime/clock/utc2tai.clk}} which maintains the clock corrections leap seconds in TEMPO2 generally.

On Unix-like systems, a \verb#crontab# can be used to ensure these files are always up-to-date. We include an example crontab in the installation notes for our code.

\subsection{Use of -tempo1 and ``prediction mode''}

TEMPO2 is designed to model pulse arrival times, but it also has a ``prediction mode'' which will generate artificial observed pulse frequencies for a pulsar of given parameters.  Use of this mode apparently requires that TEMPO2 emulate the methods of an earlier version of \verb#TEMPO#, but these differences are not relevant for 1 cm/s barycentric correction.  TEMPO2 is called from the command line as:

\begin{verbatim}
tempo2 -f 10700.par -polyco
"53991 53993 60 12 12 ctio 545000000" -tempo1
\end{verbatim}

In practice, the parameter file should be specified with a full, absolute path.  The \verb#-tempo1# flag tells TEMPO2 to emulate \verb#TEMPO1# syntax and calculation --- prediction mode using TEMPO2 methods is not documented.

The \verb#-polyco# option tells TEMPO2 to return a file containing the polynomial coefficients that describe the pulse arrival times of the pulsar.  The argument in quotation marks contains the first and last days of the results in MJD, the number of minutes over which a single set of polynomials should be valid, the order of the polynomial (12 coefficients), the maximum hour angle of the target (which we may set to 12 so as to get results for all times), the code of the observatory (from optical.dat), and the frequency of the observations in megahertz (which we have set to green light --- it is unclear to what degree this value matters as such high frequencies in TEMPO2).  Due to a quirk in TEMPO2, the first and last dates requested must be one day removed from the day for which one requires results (or else the time the user needs may not be included in the resulting output file).

The output file \verb#polyco.dat# must then be parsed to determine the observed frequency of a nominal pulsar at a given time.  For observations spanning a leap second, there is an apparent discontinuity in frequencies because UTC is discontinuous there.  For such times, we compute frequencies on either side of the leap second and interpolate.

\subsection{What TEMPO2 reports}

TEMPO2 in prediction mode calculates the observed pulse arrival times and frequencies of those pulses at an observatory on the Earth (\numeas), given the true frequency that would be observed at the SSB ($\nu^\prime$).  Since, by construction, our pulsar has no perturbers (i.e. it is our frequency standard), we have $z_{true} = 0$, so by Eqs. \ref{zb} and \ref{zmeaseq} we have that the observed frequencies \verb#F# are given by:

\begin{equation}
\verb#F# = \numeas = \frac{\nu^\prime}{(1+\zmeas)} = \nu^\prime(1+z_B)
\end{equation}

\noindent to the precision that these equations are valid descriptions of the problem.  Thus, since we have chosen $\nu^\prime = 1$ Hz, the frequencies \verb#F# reported by TEMPO2 (given by the polynomial coefficients in \verb#polyco.dat# in units of Hz) can be compared with our calculations of $z_B$ using:

\begin{equation}
z_B = \verb#F# - 1
\end{equation}

The precision of this equality describes the precision of one's barycentric correction for standard stars, compared with that computed using TEMPO2.


\end{document}